\documentclass{mn2e}

\def\lapprox{\hbox{\lower .8ex\hbox{$\,\buildrel < \over\sim\,$}}}
\def\gapprox{\hbox{\lower .8ex\hbox{$\,\buildrel > \over\sim\,$}}}

\def\apj{ApJ}
\def\aap{A\&A}

\def\apjs{ApJS}

\def\mnras{MNRAS}

\def\pasp{PASP}

\title{High proper motion white dwarfs and halo dark matter}

\author[S. Torres et al.]{Santiago Torres$^{1,2}$,
                         Enrique Garc\'\i a--Berro$^{2,3}$,
                         Andreas Burkert$^4$, and
                         Jordi Isern$^{3,5}$ \\
$^1$Departament  de  Telecomunicaci\'o  i Arquitectura  de  Computadors,
    E.U.P.  de Matar\'o, Universitat  Polit\`ecnica de Catalunya, \\ Av.
    Puig i Cadafalch 101, 08303 Matar\'o, Spain\\
$^2$Departament  de F\'\i sica Aplicada,  Universitat  Polit\`ecnica  de
    Catalunya,  Jordi Girona  Salgado s/n, M\`odul  B-4, Campus Nord, \\
    08034 Barcelona, Spain\\
$^3$Institut d'Estudis Espacials de Catalunya (C.S.I.C./U.P.C.), Edifici
    Nexus, Gran Capit\`a 2-4, 08034 Barcelona, Spain \\
$^4$Max-Planck-Institut   f\"ur   Astronomie,   Koenigstuhl   17,  69117
    Heidelberg, Germany\\
$^5$Institut de Ci\`encies de l'Espai, C.S.I.C.}

\begin{document}

\maketitle

\begin{abstract}  
The  interpretation  of the old, cool  white  dwarfs  recently  found by
Oppenheimer  et  al.  (2001)  is  still   controversial.  Whereas  these
authors  claim that they have  finally  found the elusive  ancient  halo
white dwarf population that contributes significantly to the mass budget
of the galactic halo, there have been several other  contributions  that
argue that these white dwarfs are not genuine halo members but, instead,
thick disk stars.  We show here that the  interpretation  of this sample
is  based  on  the  adopted   distances,   which  are  obtained  from  a
color--magnitude  calibration, and we demonstrate  that when the correct
distances  are used a sizeable  fraction  of these  putative  halo white
dwarfs belong indeed to the disk  population.  We also perform a maximum
likelihood  analysis of the  remaining  set of white  dwarfs and we find
that they most  likely  belong to the thick  disk  population.  However,
another  possible  explanation  is that this sample of white  dwarfs has
been  drawn  from a 1:1  mixture  of  the  halo  and  disk  white  dwarf
populations.
\end{abstract}


\section{Introduction}

White  dwarfs are the most  common  end--points  of  stellar  evolution.
Since they are long-lived and well understood  objects, they  constitute
an invaluable tool to study the evolution and structure of our Galaxy in
general and of the Galactic  halo in  particular  (Isern et al.  1998a).
Moreover,  the  discovery of  microlenses  towards the Large  Magellanic
Cloud (Alcock et al.  2000; Lasserre et al.  2001) has generated a large
controversy about the possibility that white dwarfs could be responsible
for these  microlensing  events and, thus, could  provide a  significant
contribution  to the mass budget of our Galactic  halo.  However,  white
dwarfs as viable dark matter  candidates are not free of problems, since
an excess of them would  imply as well an  overproduction  of red dwarfs
and Type II  supernovae.  In order to  overcome  this  problem  Adams \&
Laughlin (1996) proposed a non--standard  initial mass function in which
the  formation  of both low and high mass stars was  suppresed.  Besides
the lack of evidence for such biased initial mass  functions,  they also
present additional problems.  The formation of an average mass ($\sim \,
0.6\,  M_\odot$)  white dwarf is accompanied  by the injection  into the
interstellar medium of a sizeable amount of mass (typically $\sim 1.5 \,
M_\odot$) per white white dwarf.  Since in turn Type II  supernovae  are
suppressed in biased initial mass functions,  there is not enough energy
to eject this matter  into the  intergalactic  medium and a mass that is
roughly  three  times the mass of the  resulting  white  dwarf has to be
accomodated into the Galaxy (Isern et al.  1998).  Furthermore, the mass
ejected in the process of  formation  of a white dwarf is  significantly
enriched in metals (Abia et al.  2001; Gibson \& Mould  1997).  Finally,
an excess of white  dwarfs  may  translate  into an excess  of  binaries
containing  such stars.  If there are many white dwarfs in binaries then
the secondary  cannot be a red dwarf because  these would these would be
easily  detected.  Therefore,  we are then  forced to assume  that these
binaries are double degenerates, which are one of the currently proposed
scenarios  for Type Ia  supernovae.  Hence  we are  forced  to face  the
subsequent  increase of Type Ia  supernova  rates  which,  consequently,
results in an increase in the  abundances  of the  elements  of the iron
peak   (Canal,   Isern   \&   Ruiz--Lapuente   1997).   However,   other
explanations,  such as self--lensing  in the LMC (Wu 1994; Salati et al.
1999), or background  objects (Green \& Jedamzik  2002) are possible and
have not been yet totally ruled out.

The debate of whether or not white dwarfs  contribute  significantly  to
the  Galactic  halo  dark  matter  has   motivated  a  large  number  of
observational  searches  (Knox,  Hawkins  \& Hambly  1999;  Ibata et al.
1999;  Oppenheimer et al.  2001;  Majewski \& Siegel 2002, Nelson et al.
2002) and theoretical works (Reyl\'e, Robin \& Crez\'e 2001; Koopmans \&
Blandford  2002; Flynn,  Holopainen \& Holmberg 2002) and is still open.
Among the  observational  surveys perhaps the most extensive one is that
of Oppenheimer et al.  (2001) who  discovered 38 faint white dwarfs with
large  proper  motions  in  digitized   photographic   plates  from  the
SuperCOSMOS  Sky Survey.  Oppenheimer et al.  (2001)  claimed that these
white  dwarfs are indeed  halo white  dwarfs  since they have very large
tangential  velocities  (in excess of $\sim 100$ km s$^{-1}$).  Based on
this  assumption,  they  derived a space  density of 2\% of the Galactic
dark halo density, which is smaller than previous  claims (Alcock et al.
1997)  for halo  dark  matter  in the form of  $\approx  0.5\,  M_\odot$
objects, but still  significant.  However,  Reid, Sahu \& Hawley  (2001)
challenged  this  claim by noting  that the  kinematics  of these  white
dwarfs is  consistent  with the  high--velocity  tail of the thick disk.
Hansen (2001)  provided  evidence that this sample  presents a spread in
age that makes it more  likely to belong to the thick  disk  population.
Reyl\'e  et al.  (2001)  and  Flynn  et al.  (2002)  also  support  this
interpretation.  Koopmans \& Blanford (2002) find that the  contribution
of these white dwarfs to the local halo dark matter  density is smaller,
of the order of 0.8\%, which is in good agreement  with the  theoretical
results of Isern et al.  (1998b) and the  observational  findings of the
EROS team (Goldman et al.  2002).  In this paper we reexamine this issue
by making  use of a Monte  Carlo  simulator  (Garc\'\i  a--Berro  et al.
1999;  Torres et al.  1998).  The  paper is  organized  as  follows.  In
section  \S 2  we  present  the  main  properties  of  our  Monte  Carlo
simulator.  In  \S 3 we  discuss  the  effect  of  the  color--magnitude
calibration  on the  distances  of the  white  dwarfs  in the  sample of
Oppenheimer  et al.  (2001)  whereas  in \S 4 we  analyze  which  is the
probability of this sample to be drawn from a halo  population.  Finally
in \S 5 our conclusions are summarized.

\section{The model}

A full description of our Monte Carlo simulator can be found in Garc\'\i
a--Berro et al.  (1999).  Therefore we will only summarize here the most
important  inputs.  Our model includes two components:  the disk and the
stellar halo.  We start with the disk model.  Firstly,  masses and birth
times are drawn  according to a standard  initial mass  function  (Scalo
1998)  and an  exponentially  decreasing  star  formation  rate per unit
surface  area  (Bravo,  Isern \& Canal  1993;  Isern et al.  1995).  The
spatial density  distribution is obtained from a scale height law (Isern
et al.  1995)  which  varies  with time and is related  to the  velocity
distributions --- see below --- and an exponentially  decreasing surface
density in the Galactocentric distance.  The velocities of the simulated
stars are drawn from Gaussian distributions.  The Gaussian distributions
take into  account  both the  differential  rotation of the disk and the
peculiar  velocity  of the  Sun  (Dehnen  \&  Binney  1997).  The  three
components of the velocity dispersion  $(\sigma_{\rm U}, \sigma_{\rm V},
\sigma_{\rm  W})$ and the lag velocity $V_0$ are not  independent of the
scale height but,  instead,  are taken from the fit of Mihalas \& Binney
(1981) to main sequence star counts.  It is important to realize at this
point that with this  description we recover both the thick and the thin
disk populations,  and, moreover, we obtain an excellent fit to the disk
white dwarf luminosity  function (Garc\'\i  a--Berro et al.  1999).  For
the stellar halo model we adopt a  spherically  symmetric  stellar  halo
with a density profile given by the expression:

\begin{equation}
\rho=\rho_0\Big(\frac{R_\odot}{r}\Big)^\gamma
\end{equation}

\noindent where $\rho_0$ is the local density of the halo, $\gamma=3.4$,
and $R_\odot=8.5$  kpc is the  Galactocentric  distance of the sun.  The
velocity distributions are Gaussian:

\begin{equation}
f(v_r,v_{\theta},v_{\phi})=\frac{1}{(2\pi)^{3/2}}
\frac{1}{\sigma_r\sigma_t^2}\exp\left[-\frac{1}{2}\left(
\frac{v_r^2}{\sigma_r^2}+\frac{v_{\theta}^2+v_{\phi}^2}{\sigma_t^2}
\right)\right]
\end{equation}

\noindent The radial and tangential velocity  dispersions are determined
from  Markovi\'c  \&  Sommer--Larsen  (1996).  For the  radial  velocity
dispersion we have:

\begin{equation}
\sigma_r^2=\sigma_0^2+\sigma_+^2\left[\frac{1}{2}
-\frac{1}{\pi}\arctan\left(\frac{r-r_0}{l}\right)\right]
\end{equation}

\noindent where $\sigma_0=80\, {\rm km\,s^{-1}}$,  $\sigma_+=145\,  {\rm
km\,s^{-1}}$, $r_0=10.5$ kpc and $l=5.5$ kpc.  The tangential dispersion
is given by:

\begin{equation}
\sigma_t^2=\frac{1}{2}V_{\rm c}^2-\left(\frac{\gamma}{2}-1\right)\sigma_r^2+
\frac{r}{2}\frac{{\rm d}\sigma_r^2}{{\rm d}r}
\end{equation}

\noindent where 
\begin{equation}
r\frac{{\rm d}\sigma_r^2}{{\rm d}r}=
-\frac{1}{\pi}\frac{r}{l}\frac{\sigma_+^2}{1+[(r-r_0)/l]^2}
\end{equation}

\noindent For the calculations  reported here we have adopted a circular
velocity $V_{\rm c}= 220$~km/s.

The halo was assumed to be formed in an intense burst of star  formation
that  occured 14 Gyr ago and lasted  for 1 Gyr.  Regarding  the  cooling
sequences,  we adopt those of Salaris et al.  (2000)  which  incorporate
the most accurate physical inputs for the stellar interior and reproduce
the  blue  turn  due to  the  hydrogen  opacity  (Hansen  1999)  at  low
luminosities.  We use the transformations of Bessell (1986) and Blair \&
Gilmore  (1982) to convert  the colors of the  atmospheres  of Saumon \&
Jacobson  (1999) to the  photographic  colors used by Oppenheimer et al.
(2001).  Main  sequence  lifetimes  and  the  initial  mass--final  mass
relationship  for  white  dwarfs  are as in  Garc\'\i  a--Berro  et  al.
(1999).  The initial mass  function  adopted for the halo is the same as
for  the  disk  simulations  (Scalo  1998).  Finally  the  observational
selection  criteria adopted in order to draw white dwarfs from the Monte
Carlo simulated  populations  are the same as used by Oppenheimer et al.
(2001),   namely,   $0.^{\prime\prime}16\,{\rm    yr}^{-1}\le  \mu   \le
10.^{\prime\prime}0\,{\rm yr}^{-1}$ in proper motion, $16.6^{\rm mag}\le
R_{\rm 59F}\le  19.8^{\rm mag}$ in apparent  magnitude,  distances $d\le
200$ pc and 4900 square  degrees in the direction of the South  Galactic
Cap.

\section{The color--magnitude calibration}

\begin{figure}
\centering
\vspace{13cm}
\includegraphics{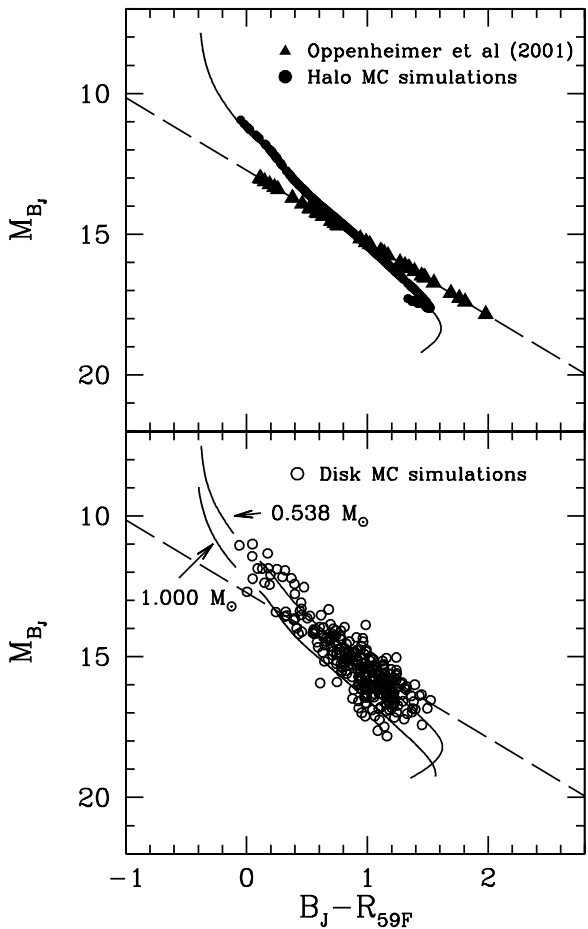}
\caption{Top   panel:  the   calibration  of  the  $M_{\rm  B_{\rm  J}}$
	 magnitude as a function the $B_{\rm J}-R_{\rm 59F}$ color index
	 used in Oppenheimer et al.  (2001) --- dashed line --- compared
	 to the cooling track of a $0.6\, M_\odot$ white dwarf --- solid
	 line --- and a typical Monte Carlo simulation of the halo white
	 dwarf  population  --- circles.  The white dwarfs in the sample
	 of  Oppenheimer  et al.  (2001) are  represented  as triangles.
	 Bottom  panel:  A typical  Monte Carlo  simulation  of the disk
	 white dwarf population.  See text for details.}
\end{figure}

Oppenheimer et al.  (2001) used the observational data of Bergeron, Ruiz
\& Leggett  (1997) to obtain an  empirical  calibration  of the  $M_{\rm
B_{\rm J}}$ magnitude from the $B_{\rm  J}-R_{\rm  59F}$ color index and
from it the distances of the white dwarfs.  In the top panel of figure 1
we show this  calibration  as a dashed  line.  The white  dwarfs  of the
sample of Oppenheimer et al.  (2001) are represented as triangles.  Also
shown in this panel is the result of a typical Monte Carlo simulation of
the halo white dwarf  population  (circles) and the cooling  track of an
otherwise  typical $0.6\,  M_\odot$  white dwarf, solid line (Salaris et
al.  2000).  As can be seen in  this  figure  there  are  two  prominent
features.  Firstly, and most  importantly,  the slope of the calibration
of Oppenheimer  et al.  (2001) is very  different  compared with that of
the Monte Carlo  simulation.  As a result,  the  distances  of the white
dwarfs  of  the  sample  of   Oppenheimer   et  al.  (2001)   have  been
underestimated  for  bright  white  dwarfs  and  overestimated  for  low
luminosity white dwarfs.  Secondly, the presence of the downturn at very
low  luminosities is clearly seen.  For the adopted age of the halo, the
turn-off  to the blue for the  Monte  Carlo  simulation  is  located  at
$B_{\rm J}-R_{\rm  59F}\simeq 1.2$ whereas for a typical $0.6\, M_\odot$
white  dwarf  is  located  at  $B_{\rm  J}-R_{\rm  59F}=1.6$.  This is a
consequence  of the adopted age of the halo.  Since the halo was assumed
to be  formed  as a burst  of star  formation,  halo  white  dwarfs  are
distributed along the color--magnitude  diagram according to their mass,
given  that  the  relation  $t_{\rm  halo}\simeq  t_{\rm   MS}(M)+t_{\rm
cool}(L,M)$  always  holds  (Isern  et al.  1998a).  Clearly,  the white
dwarfs which are beyond the turn-off in the Monte Carlo  simulation  are
massive white dwarfs, which come from massive  progenitors  with smaller
main sequence  lifetimes.  Nevertheless, the turn-offs of both the Monte
Carlo  simulation  and the  cooling  track of Salaris et al.  (2000) are
located at bluer colors than the coolest  white  dwarfs in the sample of
Oppenheimer  et al.  (2001).  Therefore,  since these  white  dwarfs are
clearly beyond both turn-offs they cannot be DA white dwarfs.

\begin{figure}
\centering
\vspace{13cm}
\includegraphics{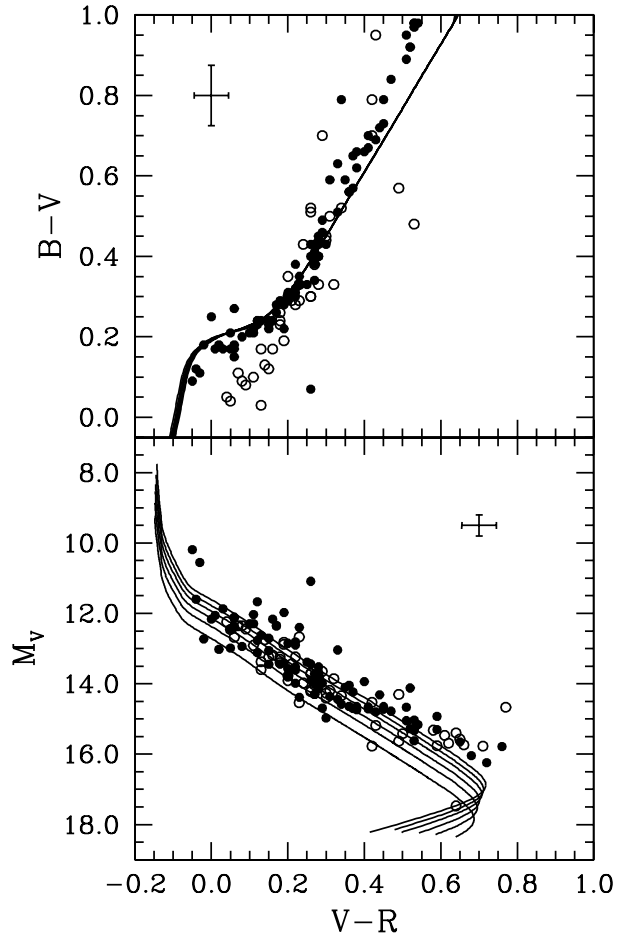}
\caption{Top panel:  color--color  diagram for the sample of Bergeron et
	 al.  (2001).  We have  chosen  this  diagram  because  in these
	 colors the dispersion of the cooling  sequences is minimum.  He
	 white dwarfs are  represented  as open symbols  whereas  filled
	 symbols  correspond  to white  dwarfs with H-rich  atmospheres.
	 The cooling tracks of Salaris et al.  (2000) for several masses
	 are  also  shown.  Bottom  panel:  comparison  of  the  cooling
	 tracks  of  Salaris  et al.  (2000)  with the  color--magnitude
	 diagram of the same sample of old, cool disk white dwarfs.  The
	 masses of the  cooling  tracks  are, from top to bottom,  0.54,
	 0.61,  0.68,  0.87  and 1.0  $M_\odot$,  respectively.  In both
	 panels the typical error bars are shown.}
\end{figure}

Now the following question arises:  which process is responsible for the
different slopes in the color--magnitude  calibration?  An idea would be
that, in  principle,  this  difference  could be ascribed  solely to the
different  physics  of the  adopted  envelopes.  The  cooling  sequences
adopted  in  this  work  are  those  of  Salaris  et  al.  (2000)  which
incorporate the most up-to-date  atmospheres  (Saumon \& Jacobson 1999).
In figure 2 we compare the cooling tracks of Salaris et al.  (2000) with
the  observational  data of Bergeron  et al.  (2001).  As it can be seen
our  cooling  sequences  compare  very  favourably  with  the  available
observational data both in the color--color and in the  color--magnitude
diagram.  For instance, most of the  overluminous  white dwarfs  located
above the  theoretical  cooling track of the $0.54\,  M_\odot$ model are
unresolved  binaries and, as discussed in Bergeron et al.  (2001), their
luminosity  comes from the  contribution  of two otherwise  normal white
dwarfs.  In the  color--color  diagram the agreement is also  excellent,
especially  for  $V-R<0.4$.  For $V-R>0.4$ a slight  departure  from the
observational  data is  observed,  but always  within the  observational
error bars.  Again, as discussed  in Bergeron et al.  (2001),  this is a
common drawback of all theoretical  models and can be explained in terms
of a missing  opacity  source  near the $B$ filter in the pure  hydrogen
models, most likely due to a  pseudocontinuum  opacity  originating from
the Lyman edge.  Therefore we conclude that our cooling sequences are in
good  agreement  with the  observational  data for old, cool disk  white
dwarfs.  Note, however, the excess of  overluminous  white dwarfs at the
red end of the  cooling  sequences.  We will  come  back to  this  issue
later.

Although this could be indeed one of the reasons there is still  another
possibility,  namely, that the calibration of Oppenheimer et al.  (2001)
is not  appropiate  for the halo  white  dwarf  population.  The  reader
should take into account  that  Oppenheimer  et al.  (2001)  derived the
above  mentioned  calibration  using a sample of cool disk white dwarfs,
namely with $M_{\rm  V}>12$.  Note that for $M_{\rm  V}\sim 12$ there is
an abrupt change in the slope of the cooling tracks.  In this regard, in
the bottom panel of figure 1 we show the result of a typical Monte Carlo
simulation  of the disk white dwarf  population  (open  circles) and the
calibration used by Oppenheimer et al.  (2001).  Since they obtained the
calibration using white dwarfs with known parallaxes for which the error
in the  parallax  determination  was  smaller  than 30\% we have added a
conservative  gaussian  error of 20\% for the  parallaxes  of the  white
dwarfs in this sample.  Additionally a 10\% error in the color index has
also been  added.  There is as well  another  spread in the  photometric
calibration which comes from the very different star formation histories
of both populations.  Indeed the disk white dwarf population is obtained
from a smoothly varying star formation rate which produces massive white
dwarfs  almost  continously  as a  consequence  of the very  small  main
sequence   lifetimes  of  their   progenitors   whereas,  as  previously
discussed, the halo white dwarf  population is distributed  according to
the mass along the  cooling  track of a typical  $0.6\,  M_\odot$  white
dwarf.  The mass spread is clearly seen in the bottom panel of figure 1,
where the cooling  tracks of Salaris et al.  (2000) for a 0.538 and $1.0
\,  M_\odot$  white  dwarfs  are  shown.  All  these  effects  force the
distribution  of disk white dwarfs to have a  significant  spread in the
color--magnitude  diagram.  Moreover,  as can be seen in this panel, the
slope of the  calibration of  Oppenheimer  et al.  (2001) could be valid
for a  randomly  selected  sample of cool  white  dwarfs,  that is white
dwarfs with colors $B_{\rm J}-R_{\rm 59F}\gapprox 0.5$ or, equivalently,
$M_{\rm  V}\gapprox  12$.  In fact, since  Bergeron  et al.  (1997) were
selecting  cool white  dwarfs,  namely with  $M_{\rm  V}\gapprox  12$, a
shallower slope of photometric  calibration would not be very surprising
given the  observational  errors.  In order to check  this and to make a
more quantitative statement we have randomly selected from our simulated
samples 80 subsets of 100 white dwarfs, which is the typical size of the
sample of Bergeron et al.  (1997), with $M_{\rm V}>12$.  For each of the
subsets we have computed the slope of the  color--magnitude  calibration
and its standard  deviation.  We obtain a mean slope and a mean standard
deviation  of $3.18 \pm 0.25$ for the  $M_{\rm  B_{\rm  J}}$  versus the
$B_{\rm  J}-R_{\rm  59F}$  calibration.  Here we have  used for the mean
standard  deviation the ensemble  average of the individual  dispersions
for each one of the  subsets.  This value has to be  compared  with that
adopted by  Oppenheimer  et al.  (2001),  namely  $2.58$, which still is
slightly beyond the $2\sigma$ confidence interval.  Hence, although this
is a possible  explanation of the discrepancy in the slopes there may be
another effect at the root of this discrepancy.

\begin{figure}
\centering
\vspace{13cm}
\includegraphics{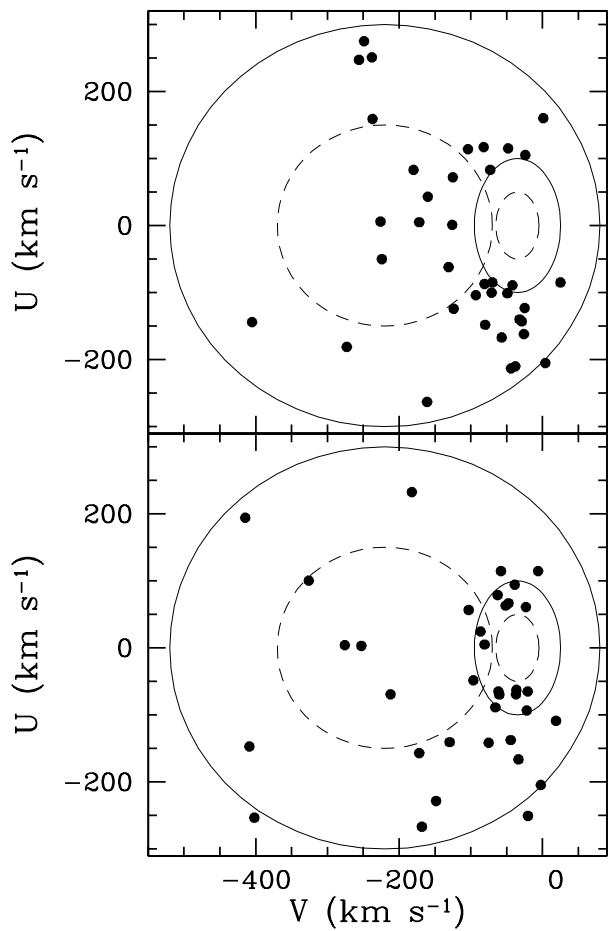}
\caption{Panels showing the tangential velocities of the white dwarfs of
	 the sample of Oppenheimer et al.  (2001).  In the top panel the
	 velocities  obtained  using the distances  obtained  from their
	 calibration   are  shown,  whereas  in  the  bottom  panel  the
	 distances  obtained  in this work have been used.  We have also
	 assumed null radial velocity as Oppenheimer et al.  (2001) did.
	 See text for further details.}
\end{figure}

Indeed, there is another subtle effect that should be taken into account
in  analyzing  the  color--magnitude   calibration.  Note  that  in  the
color--magnitude  of  figure 2 the blue  portion  of this  diagram  (say
$V-R<0.4$) is more populated  that the red --- and, hence, cool --- part
of the diagram.  The observational disk white dwarf luminosity  function
shows a  monotonic  increase  all  the  way to  $M_{\rm  V}  \simeq  15$
(Leggett, Ruiz \& Bergeron, 1998), and a sharp drop at $M_{\rm V} \simeq
16$ as a consequence  of the finite age of the disk.  Note that the blue
turn is not visible in the disk white dwarf  population  since it occurs
at even fainter  luminosities  ($M_{\rm V}\simeq 18$).  Hence, we should
expect an increasing  number of white dwarfs at red colors.  This is not
what it is  observationally  found and, in fact,  there is an  otherwise
natural selection effect against low luminosity white dwarfs.  Moreover,
figure 2 clearly shows that low-mass white dwarfs, those with $M<0.54 \,
M_\odot$,  are  more  abundant  at the red  end of the  color--magnitude
diagram.  As noted above  these  low-mass  white  dwarfs are  members of
unresolved  binaries, and this  explains why they are  overluminous.  In
turn, the fact that these white  dwarfs are  overluminous  explains  why
they are more abundant at low luminosities.  But this observational bias
strongly affects the slope of the  color--magnitude  calibration, making
it shallower.  Moreover, from the theoretical point of view there are as
well  compelling  evidences  to  exclude  these  white  dwarfs  from the
color--magnitude  calibration since single low-mass white dwarfs have He
cores and their  progenitors  have not had enough time to evolve off the
main  sequence.  In any case, the  important  point  here is that  these
overluminous   white   dwarfs   dominate   the   red   portion   of  the
color--magnitude  diagram and, hence, the color--magnitude  calibration.
In  order  to make  this  argument  quantitative  we have  proceeded  as
follows.  We  eliminate  from the sample of Bergeron  et al.  (2001) all
white dwarfs with masses smaller than $0.54\, M_\odot$, because they are
suspected to be unresolved binaries.  After that we compute a linear fit
to the empirical  cooling  sequence for $M_{\rm  V}>12$.  We obtain that
the  slope of the  linear  fit is  $2.95  \pm  0.18$,  which  is in good
agreement with the result obtained from the Monte Carlo simulations.  In
summary, there are clear evidences from both the  observational  and the
theoretical   point  of  view  to  adopt  a   steeper   color--magnitude
calibration  in  accordance  with  the  theoretical   models.  Thus,  we
conclude that the  distances  derived for the white dwarfs in the sample
of  Oppenheimer  et al.  (2001) should be  recomputed  using the correct
cooling tracks.

The basic  argument  used by  Oppenheimer  et al.  (2001) to claim  that
their sample is representative of an ancient halo white dwarf population
was that these white dwarfs have very large tangential velocities.  This
result is  sensitive  to the  adopted  distances.  Moreover,  since  the
distances  of  bright  white  dwarfs  have  been  overestimated  and the
distances of dim white dwarfs have been underestimated it is not evident
how the  color--magnitude  calibration  affects the  derived  tangential
velocities.  This  is  assesed   in  figure  3,  where  the   tangential
velocities  of the white  dwarfs of the  sample  of  Oppenheimer  et al.
(2001) are shown.  We followed  exactly  the same  procedure  they used.
That is, we have  assumed  null  radial  velocity.  In the top  panel of
figure 3 the velocities  obtained using the distances  computed from the
calibration  of  Oppenheimer  et al.  (2001) are shown,  whereas  in the
bottom panel the distances obtained in this work have been used.  In the
bottom panel of figure 3 the white dwarfs  which are located  beyond the
blue turn-off in figure 1 have been removed since our cooling  sequences
are  not  able  to  reproduce  their  position  in the  color--magnitude
diagram.  Also shown in this figure are the velocity  ellipsoids for the
disk and the halo (at 1$\sigma$ and 2$\sigma$).  The velocity ellipsoids
for the halo  are  centered  at  $(U,V)=(0,-220)$~km/s.  The  radius  at
$1\sigma$  is given  by  $\sigma_{\rm  U}=\sigma_{\rm  V} =  V_{\rm  c}/
\sqrt{2}$.  The velocity ellipsoids for the disk are centered at $(U,V)=
(0,-35)$~km/s.  The axis at $1\sigma$ are $(\sigma_{\rm  U}, \sigma_{\rm
V})= (50,30)$~km/s (Dehnen \& Binney, 1998).

As can be seen in figure 3 the resulting tangential  velocities are such
that a  significant  fraction  of the  white  dwarfs  of the  sample  of
Oppenheimer  et al.  (2001) move inside the  velocity  ellipsoid  of the
disk.  Therefore, and following the same criterion  used by  Oppenheimer
et al.  (2001)  these white  dwarfs are not  genuine  halo  members  and
should be dropped from further analysis.

\section{Maximum likelihood analysis of the sample}

Now we  concentrate  our  efforts  on  performing  a maximum  likelihood
analysis  of the  potential  halo white  dwarf  candidates  found in the
sample  of  Oppenheimer  et al.  (2001).  In  order  to do so we use the
following  procedure.  First it should be noted that  Oppenheimer et al.
(2001)  disregarded all white dwarfs situated  inside the $2\sigma$ disk
contour of figure 3.  As previously  stated, we follow  exactly the same
criterion.  There are 23 white  dwarfs for which the  distances  derived
here are beyond the $2\sigma$  contour of the disk  velocity  ellipsoid.
These white dwarfs are  represented as large filled circles in figure 4.
Of these  white  dwarfs  there are 8 which  are  located  in the  region
between the $2\sigma$  and  $4\sigma$  contours  (shown in figure 4 as a
long  dashed  line) of the disk  population.  That is, there are 8 white
dwarfs  located in what we can call the most  extreme  tail of the thick
disk  distribution.  We generate  Monte Carlo  simulations  for both the
disk and the halo, with exactly the same  restrictions  in magnitude and
proper motion adopted by  Oppenheimer  et al.  (2001) and located in the
same  region of the sky.  These  simulations  are  shown in  figure 4 as
small open  circles.  The  number of stars in both  simulations  is very
large (of the order of $\sim 10^4$) but, for the sake of clarity, only a
small fraction of randomly selected white dwarfs has been represented in
these diagrams.  From these  simulations we extract a subset of 23 white
dwarfs.  Then we count how many white  dwarfs of this subset are located
in the region between the $2\sigma$ and $4\sigma$ contours.  Let us call
this number $n$.  We repeat the process  iteratively  many times, of the
order of $N=10^6$,  until  significant  statistics  are achieved  and we
compute the number of times $N_n$ that we find $n$ white  dwarfs in this
region.  The  probability  of $n$ stars to be located in this  region of
the diagram  (between the  $2\sigma$  and  $4\sigma$  contours)  is then
$P=N_n/N$.  We compute this probability for both the halo simulation and
the disk simulation.

\begin{figure}
\centering
\vspace{13cm}
\includegraphics{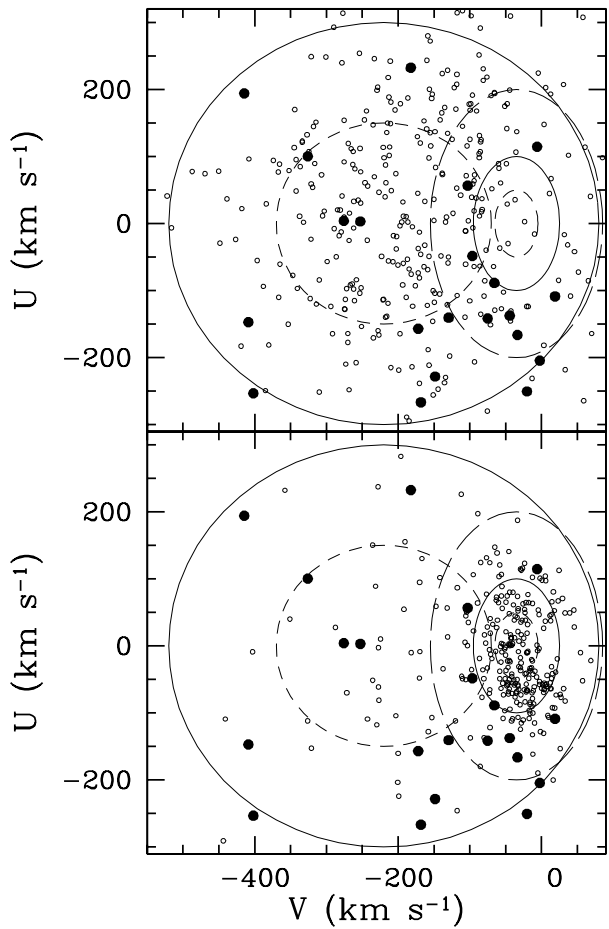}
\caption{The  distribution of tangential  velocities of the white dwarfs
	 of the sample of Oppenheimer  et al.  (2001),  filled  symbols,
	 compared to the  results of a typical  Monte  Carlo  simulation
	 (open  symbols)  of the halo --- top panel  --- and of the disk
	 --- bottom panel.  We have adopted our revised distances to the
	 the white  dwarfs of  Oppenheimer  et al.  (2001) in  computing
	 their velocities.}
\end{figure}

\begin{figure}
\centering
\vspace{13cm}
\includegraphics{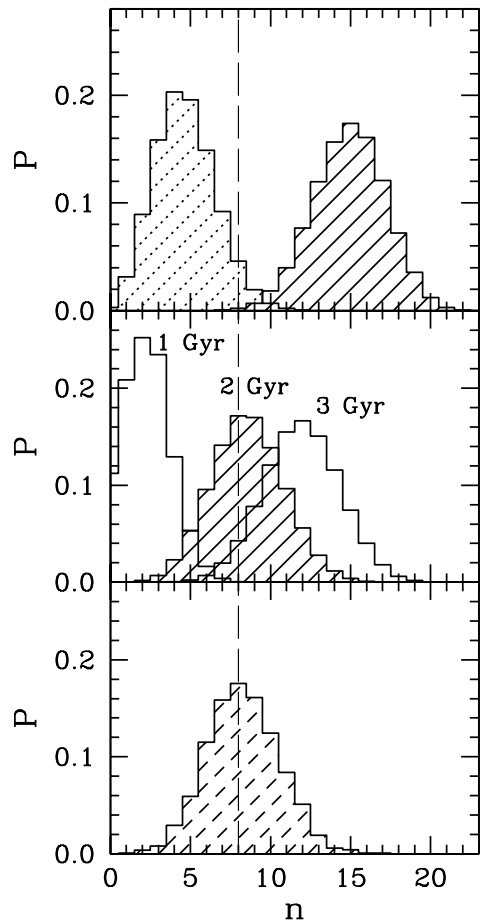}
\caption{Distribution of normalized  probabilities that the white dwarfs
	 of the sample of Oppenheimer  et al.  (2001) are drawn from the
	 disk and halo simulations  shown in figure 4.  The left diagram
	 of the top panel corresponds to the halo population whereas the
	 right diagram  corresponds  to the disk  simulation.  The three
	 histograms of the central panel  correspond  to the stars which
	 were born in the very  early  stages of the life of the disk of
	 our Galaxy (1, 2 and 3 Gyr,  respectively).  Finally the bottom
	 panel displays the normalized  probability  distribution  for a
	 1:1 mixture of halo and disk stars.}
\end{figure}

\begin{figure}
\centering
\vspace{13cm}
\includegraphics{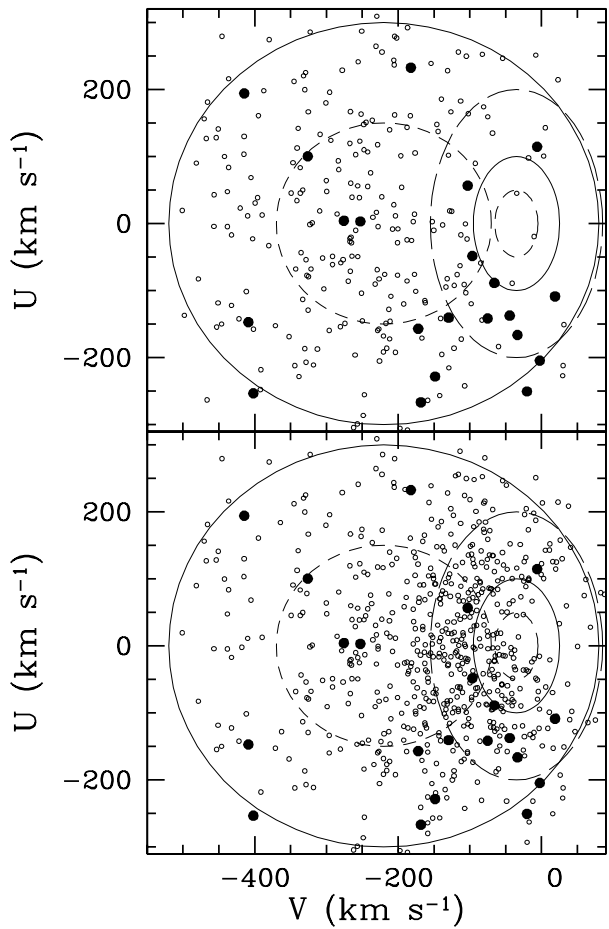}
\caption{Panel  showing the distribution of tangential velocities of the
	 white  dwarfs  of the  sample  of  Oppenheimer  et al.  (2001),
	 filled  symbols,  compared to the disk white  dwarf  stars with
	 birth times  smaller than 1 Gyr --- top panel --- and 2 Gyr ---
	 bottom panel --- obtained in a typical Monte Carlo simulation.}
\end{figure}

The probabilities  computed with the above explained procedure are shown
in the top  panel  of  figure  5.  As can be seen  in  this  panel  both
distributions  of  normalized  probabilities  are  Gaussian  to  a  good
approximation.  The  distribution  of  probabilities  for the halo (left
histogram) is centered at $n=5$, whereas the corresponding  distribution
for the disk is centered at $n=15$.  Their full  widths at half  maximum
are,  respectively,  $\simeq  4$ and  $\simeq  5$.  The  number of white
dwarfs of the  sample  of  Oppenheimer  et al.  (2001)  located  in this
region  is  marked  as a thin  dashed  line.  It is  thus  difficult  to
ascertain  whether  or not  the  sample  of  Oppenheimer  et al.  (2001)
belongs to the halo  population  or to the disk  population.  In fact it
would be possible that this sample contains stars from the tails of both
populations.  This is  important  since, in contrast  with what  happens
with main sequence stars for which their metallicity is a good indicator
of the  population to which they belong, for the case of white dwarfs we
do not have any way to  ascertain  whether a white dwarf  belongs to the
thick or to the thin disk  population,  except for its kinematics.  Most
important,  this result taken at face value  implies  that the sample of
Oppenheimer  et al.  (2001)  cannot  be  uniquely  assigned  at the 95\%
confidence  level to either  of the two  populations  {\sl as a  whole}.
Moreover  since the  fraction of thick disk stars is small in a randomly
selected  sample  of  disk  white  dwarfs  it is not  obvious  from  the
simulations presented here that the sample of Oppenheimer et al.  (2001)
belongs  to the thick  disk.  Since our model for the disk  white  dwarf
population  recovers  naturally  both the thick and the thin disk  white
dwarf  populations  as a function  of the birth  time of the  progenitor
stars, we have  binned the stars as a function  of their age.  The stars
belonging  to the thick  disk  population  are those  with  birth  times
smaller  than  say  $\simeq  2$  Gyr.  The  resulting  distributions  of
velocities   are  shown  in  figure  6,  where  the  white  dwarfs  with
progenitors with birth times smaller than 1 Gyr and 2 Gyr are shown (top
and bottom panel, respectively).

As can be seen in this  figure  the stars  which  were  born in the very
early stages of the life of our Galaxy have on average larger tangential
velocities  than the whole white dwarf  population, as expected.  Now we
perform  the same  probability  analysis  for these  subsets of the disk
white population.  The resulting  probability  distribution are shown in
the  midlde  panel of  figure 5 for 1, 2 and 3 Gyr.  Each  histogram  is
labeled with the corresponding  age.  Obviously the most probable number
of white dwarfs found in the region  between the 2$\sigma$ and 4$\sigma$
contours of the disk  decreases as the  considered  mean age  decreases.
However,  as clearly  seen in this panel  thick  disk  stars are able to
reproduce the number of stars found in this region.  In  particular,  if
we adopt an age cut of 2 Gyr the  number of white  dwarfs in the  region
between the $2\sigma$ and $4\sigma$ disk contours is nicely  reproduced.

However,  there is yet another  possibility,  namely  that the sample of
Oppenheimer et al.  (2001) corresponds to a randomly selected mixture of
both the halo and the  disk  populations  shown  in  figure  4.  This is
assesed in the  bottom  panel of figure 5 where we show the  probability
distribution for such a mixture of both disk and halo white white dwarfs
with  equal  proportions.  As it can be seen  there the  probability  of
finding eight white dwarfs in the above mentioned  region is maximum for
such a fraction.  Therefore,  it is quite likely as well that the sample
of  Oppenheimer  et al.  (2001) would contain  white dwarfs  coming from
both populations  (thick disk and halo) and that the respective ratio is
1:1.

\begin{table}
\caption{Number density of the halo white dwarf population.}
\centering
\begin{tabular}{ll}
\hline
\hline
Author & $n$ (pc$^{-3}$) \\
\hline
Oppenheimer et al. (2001) & $2.2\cdot 10^{-4}$  \\
Torres et al. (1998) &$1.2\cdot 10^{-5}$ \\
Gould, Flynn \& Bahcall (1998) & $2.2\cdot 10^{-5}$ \\
This work & $3.1\cdot 10^{-5}$ \\
\hline
\hline
\end{tabular}
\end{table}

Finally, we have computed the number density of halo white dwarfs of the
sample of Oppenheimer  et al.  (2001) with the new distances  derived in
this work and compared it with previous  works, as shown in Table 1.  In
doing so we have used the $V_{\rm max}^{-1}$ method (Schmidt 1968).  The
derived  number  density  of  this  sample  is  $n=6.2  \cdot   10^{-5}$
pc$^{-3}$.  According  to the  previous  discussion  this  density is an
upper limit to the density of halo white dwarfs.  This number  should be
compared  with the  density  originally  derived by  Oppenheimer  et al.
(2001), which is $n=2.2 \cdot 10^{-4}$  pc$^{-3}$,  which is a factor of
3.5 larger, with the density  derived using a neural network to identify
possible halo candidates by Torres et al.  (1998), which is $n=1.2 \cdot
10^{-5}$  pc$^{-3}$,  and with the  density  derived by Gould,  Flynn \&
Bahcall  (1998) using  subdwarf  stars, which is $n=2.2  \cdot  10^{-5}$
pc$^{-3}$.  Clearly  the local  density  derived in this work is in good
agreement  with  previous  independent  determinations.  Moreover, if we
assume  that only one out two white  dwarfs is a genuine  member  of the
halo  white  dwarf   population,   as  suggested   by  our  Monte  Carlo
simulations,   we  derive  a  number  density  of  $3.1  \cdot  10^{-5}$
pc$^{-3}$,  which is very close to the  number  density  of Gould et al.
(1998).

\section{Conclusions}

We have presented evidence that the distances of the white dwarfs in the
sample of Oppenheimer et al.  (2001) have not been correctly determined.
The ultimate reason of this is that the authors used a calibration which
is not  appropriate  for the  halo  white  dwarf  population.  Once  the
correct  calibration  is adopted it turns out that the  distances to the
most  luminous  white  dwarfs in the  sample  have been  underestimated,
whereas the distances to the white dwarfs with small  luminosities  have
been  overestimated.  We have also found that some  white  dwarfs in the
sample cannot have hydrogen dominated  atmospheres, since their position
in  the   color--magnitude   diagram  is  beyond  the   turn-off.  As  a
consequence, once the corrected distances are taken into account, a good
fraction of these putative halo white dwarfs have significantly  smaller
tangential  velocities  and can be  safely  discarded  as  genuine  halo
members.

The remaining  fraction of the sample of Oppenheimer  et al.  (2001) has
been analyzed using our Monte Carlo  simulator.  We have computed  Monte
Carlo models for the disk and the halo populations.  The disk simulation
naturally  recovers both the thin and the thick disk  populations.  Then
we  have  computed  the  probability  of the  stars  of  the  sample  of
Oppenheimer  et al.  (2001) to belong to a randomly  selected  sample of
both halo or disk white dwarfs.  Our results  indicate  that this subset
of the sample of Oppenheimer et al.  (2001) does not belong  exclusively
to either the halo or the disk population at the 95\% confidence  level.
Regarding  the disk  population  {\sl as a whole} our  results  were not
conclusive  because  of the  small  fraction  of thick  disk  stars in a
typical Monte Carlo simulation.  However once the stars with small birth
times ($\lapprox$ 2 Gyr),  corresponding to the thick disk, are selected
we find that the number of stars in the  sample  nicely  reproduces  the
values  found by  Oppenheimer  et al.  (2001),  in  agreement  with  the
results of Flynn et al.  (2002)  and  Reyl\'e  et al.  (2001).  There is
yet another possibility which has not been previously  explored.  Namely
that the sample of Oppenheimer et al.  (2001) is drawn from a mixture of
both the halo and the (thick)  disk  populations.  We have found that in
this case the  probability  is  maximum  for a 1:1 to  ratio.  Hence, we
conclude that the claim by Oppenheimer et al.  (2001) that, finally, the
elusive halo white dwarf  population has been found should be taken with
caution and more  observational  searches and theoretical work are still
needed.  Finally we have  re-derived,  using the  distances  obtained in
this work, the number  density of halo  white  dwarfs  predicted  by the
sample of  Oppenheimer  et al.  (2001).  We have found that a safe upper
limit to this density is $n=6.2 \cdot 10^{-5}$ pc$^{-3}$,  assuming that
{\sl all} the white dwarfs found by Oppenheimer  et al.  (2001) are true
halo white dwarfs.  If, as suggested by our  simulations, we assume that
only half of these  stars  are  genuine  halo  members  we find a number
density of $3.1 \cdot  10^{-5}$  pc$^{-3}$,  which is in good  agreement
with previous independent determinations.

\vspace{1cm}

\noindent {\sl  Acknowledgements.}  This work has been  supported by the
DGES grant  PB98--1183--C03--02, by the MCYT grant AYA2000--1785, by the
MCYT/DAAD grant  HA2000--0038 and by the CIRIT grants  1995SGR-0602  and
2000ACES-00017.  We would like to acknowledge the advise of Nigel Hambly
in  transforming  our cooling  sequences to the  appropiate  photometric
passbands.  We also would like to acknowledge the very valuable comments
of our referee, Chris Flynn, which greatly improved the original version
of the manuscript.

\end{document}